\documentclass[prd,epsf,aps,twocolumn]{revtex4}
\usepackage{amsmath}
\usepackage{amssymb}
\usepackage{eucal}
\usepackage[all]{xy}
\input xypic
\numberwithin{equation}{section}
\begin{document}
\title{Novel Geometrical Models of Relativistic Stars\\
       III. The Point Particle Idealization}
\author{P.~P.~Fiziev\footnote{ E-mail:\,\, fiziev@phys.uni-sofia.bg} }
\affiliation{Department of Theoretical Physics, Faculty of
Physics, Sofia University, 5 James Bourchier Boulevard,
Sofia~1164, Bulgaria.\\and\\ Joint Institute of Nuclear Research,
Dubna, Russia. }
\begin{abstract}
We describe a novel class of geometrical models of relativistic
stars. Our approach to the static spherically symmetric solutions
of Einstein equations is based on a careful physical analysis of
radial gauge conditions. It brings us to a two parameter family of
relativistic stars without stiff functional dependence between the
stelar radius and stelar mass.

As a result, a point particle idealization -- a limiting case of
bodies with finite dimension, becomes possible in GR, much like in
Newtonian gravity. We devote this article to detailed mathematical
study of this limit.

\noindent{PACS number(s): 04.20.Cv, 04.20.Jb, 04.20.Dw}
\end{abstract}
%
\sloppy
\newcommand{\lfrac}[2]{{#1}/{#2}}
\newcommand{\sfrac}[2]{{\small \hbox{${\frac {#1} {#2}}$}}}
\newcommand{\ben}{\begin{eqnarray}}
\newcommand{\een}{\end{eqnarray}}
\newcommand{\la}{\label}
\maketitle
%
\section{Introduction}

This is the third one of the series of articles, in which we
describe novel geometrical models of general relativistic stars
(GRS). The preliminary knowledge of the first two ones is highly
recommended for the right understanding of the present article.
There one can find the basic principles, equations, notations and
general results for construction of the new models
\cite{F0409456}, as well as the detailed description of the
simplest model of this type -- incompressible GRS \cite{F0409458}.

In the third article we describe the point particle limit of
bodies of finite dimension in GR. Thus we obtain a new ground for
the massive point particle solutions to Einstein equations (EE),
described earlier in \cite{F03}.

At present the vast majority of relativists do not accept the
consideration of point particles in GR as an incompatible with EE
idealization. There are different reasons: some doubts about
consistency of the theory of mathematical distributions (like
Dirac delta function $\delta({\bf r})$) with the nonlinear
character of EE; the understanding of the drastic change of
geometry of the Riemannian space-time
$\mathbb{M}^{(1,3)}\{g_{\mu\nu}\}$ in vicinity of a point with
infinite concentration of energy in it (this problem was for the
first time discussed by Brillouin \cite{Brillouin}), etc.

On the other hand, it is obvious that in Nature very distant
objects like stars look like "points" of finite mass and finite
luminosity. This fact has a proper mathematical description in
Newton theory of gravity in the language of mathematical
distributions. A formal mathematical problem is to find the
corresponding idealized treatment of such objects in GR, as well,
but up to recently no reasonable approach was known.

In articles \cite{F03} we showed that a correct mathematical
solutions to the  EE with $\delta({\bf r})$ term in the rhs do
exist. Such solutions describe a two parameter family of
analytical space-times $\mathbb{M}^{(1,3)}\{g_{\mu\nu}\}$ with
specific strong singularity at the place of the massive point
source with bare mechanical mass $M_0>0$ and Keplerian mass
$M<M_0$.

The price, one is forced to pay for such enlargement of standard
GR framework, is:

1) To consider the corresponding metric coefficients
$g_{\mu\nu}(x)$ as functions of class ${\cal C}^0$ of the
coordinates $x$. Some of them are to have a fixed finite jumps in
their first derivatives at the place of the point source,
necessary to reproduce by the Einstein tensor $G_\mu^\nu$ the
$\delta({\bf r})$ term  of the energy-momentum tensor $T_\mu^\nu$
in the rhs of EE.

2) To accept the unusual geometry of space-time around the matter
point in GR, which appeared at first in the original Schwarzschild
article \cite{Schwarzschild1}. It has been discussed by Brillouin
\cite{Brillouin} as early as in 1923. This geometry is essentially
different from the geometry around space-time points with finite
energy density in them.

It turns out that in perfect accord with Dirac intuition
\cite{Dirac}, the presence of massive point matter source in rhs
of EE ultimately forces one to consider the standard Hilbert gauge
(HG) form of the solution outside the source:
\ben g_{tt}(\rho)\!=\!1\!-\!{{\rho_G}/{\rho}},\,\,\,
g_{\rho\rho}(\rho) = -1/g_{tt}(\rho)\la{HGmetric}\een
{\em only} on the physical interval of the luminosity variable
$\rho\in (\rho_0,\infty)$, where
\ben
    \rho_0 = 2 M /(1-\varrho^2)> \rho_G\,.\la{rho_0}
\een
Here $\rho_G = 2 M$ is the Schwarzschild radius, $\varrho=M/M_0\in
(0,1)$ is the ratio of the Keplerian mass $M$ and the bare
mechanical mass $M_0$ of the matter point. This ratio describes
the gravitational defect of the mass of the point particle,
introduced in \cite{F03}, in scale invariant way \cite{F0409456}.

In the standard approach to GRS \cite{books} a consideration of
the limit $R_* \to 0$ of the geometrical radius $R_*$, for a fixed
stelar mass $m_*$, is impossible. Indeed, in this approach the
extra condition $\rho_C(m_*,R_*)=0$ on the central value of
Hilbert luminosity variable $\rho$ yields a stiff mass-radii
relation $m_*=m_*(R_*)$ \cite{F0409456}. Because of this relation,
it becomes impossible to vary the radius $R_*$, and, at the same
time, to prescribe a constant value to the mass $m_*$.

Here we shall show that our generalized treatment of the GRS with
Keplerian mass $m_*$ and proper mass $m_{0*}$, described in
\cite{F0409456, F0409458}, allows a correct limit  $R_*\to 0$ of
the stelar radius $R_*$ under one of the following two extra
conditions:

i) Fixed Keplerian mass  $m_*$;

ii) Fixed proper mass $m_{0*}$.

Thus, much like as in Newtonian theory of gravity, where these two
masses are identical, in GR one is able to obtain a natural
massive-point-particle idealization of objects of finite
dimension.

\section{Point Particle Limit in Newton Theory of Stars}

Let us first consider the theory of stars in Newton gravity. To be
specific, we will examine here the most well known polytropic
stars with EOS:
\ben
p=K\varepsilon^\gamma,\,\,\,\hbox{or}\,\,\,w:=p/\varepsilon=K\varepsilon^{\gamma-1},
\,\,\, \gamma\geq 6/5.\la{polytropic}\een
See, for example, the book by S.~Weinberg in \cite{books}.

In this case the non-relativistic energy distribution in the star
is described by the function

\ben \varepsilon^{NR}(r)\!=\!\begin{cases} {{m_*}\over{4\pi
r_*^3}}{{\xi_*}\over{|\theta^\prime(\xi_*)|}}
\left(\!\theta\big(\xi_*{{r}\over{r_*}}\big)\right)^{1\over{\gamma-1}}\!\!,
\hskip -.2truecm&\text{if $r\in [0,r_*]$};\,\,\\
0, & \text{if $r\,\overline\in\,\in [0,r_*],$}
\end{cases}\hskip .4truecm\la{NRvarepsilon}\een
where $r\equiv \rho$ is the standard radial variable in the
Euclidean 3D space, $\theta(\xi)=const\times p/\varepsilon=const
\times w$ is the Lane-Emden function, i.e. the solution of the
Cauchy problem
\ben \xi^{-2}{d\over{d\xi}}\left(\xi^2
{{d\theta}\over{d\xi}}\right)\!+\!\theta^{1/(\gamma\!-\!1)}\!=\!0,\nonumber\\
\theta(0)\!=\!1,\, d\theta/d\xi(0)\!=\!0;\la{LEEq}\een
$\xi:=\left(\varepsilon^{NR}(0)\right)^{2-\gamma}\sqrt{4\pi(\gamma-1)/K}\,r$,
$\xi_*\!=\!\xi_*(\gamma)$ is the minimal positive root of the
equation $\theta(\xi_*)\!=\!0$ and $r/r_*\!=\!\xi/\xi_*$. Then the
mass of the star is
\ben m_*= \int_0^{r_*}\!\!dr\,4\pi r^2 \varepsilon^{NR}(r).
\la{massN}\een

Let us consider the function
\ben F(\xi;\xi_*,0)=\begin{cases}
{{\xi^2}\over{\xi_*^2|\theta^\prime(\xi_*)|}}
\big(\theta(\xi)\big)^{1\over{\gamma-1}}\!\!,
\hskip -.2truecm&\text{if $\xi\in [0,\xi_*]$};\,\,\\
0, & \text{if $\xi\,\overline\in\,[0,\xi_*].$}
\end{cases}\hskip .4truecm\la{FN}\een
One easily obtains from Eq. (\ref{LEEq}):
\ben \int_{-\infty}^{\infty} F(\xi;\xi_*,0)d\xi=
\int_{0}^{\xi_*}F(\xi;\xi_*,0)d\xi=1. \la{F1}\een

Now we apply the Basic Lemma, proved in the Appendix, using
$\epsilon=r_*/\xi_*$ and values $x_0=x_1=0$, $x_2=\xi_*$,
$(x-x_0)/\epsilon={{\xi_*}\over{r_*}} r$. Then in the limit
$r_*\to 0$, $m_*=const$, $\gamma=const$ we have:
\ben  4\pi r^2 \varepsilon^{NR}(r)=m_*{{\xi_*}\over{r_*}}
F\left({{\xi_*}\over{r_*}}r;\xi_*,0\right) \rightharpoondown
m_*\delta(r).\,\,\,\la{NRlimit}\een

This way we obtain the massive point particle idealization of
bodies of finite dimension in Newtonian gravity
\footnote{According to our general scheme, in present series of
articles \cite{F0409456,F0409458}, as well as in \cite{F03}, we
are considering the static spherically symmetric solutions of EE
as one-dimensional problem. If one wishes to return back to the 3D
form of the corresponding results, one has to replace the 1D Dirac
$\delta$-function $\delta(r)$ with 3D one: $\delta^{(3)}({\bf
r})$. Formally, the last corresponds to the expression ${{1}\over
{4\pi r^2}}\delta(r)$ in our 1D considerations. As a result, in
full 3D notations we obtain
$$\varepsilon ({\bf r}) \rightharpoondown m_*
\delta^{(3)}({\bf r})$$ both in Newtonian and in GR theory of
gravity.}.

As expected, during the limiting process $r_*\to 0$ , $m_*=const$,
$\gamma=const$ the central energy density
\ben \varepsilon^{NR}_C=\varepsilon^{NR}(0)= {{m_*}\over{4\pi
r_*^3}}\, {{\xi_*}\over{|\theta^\prime(\xi_*)|}}\to\infty
\la{NRvarepsilon_C} \een
diverges as $1/r_*^3$. The limiting value of the central pressure
diverges stronger than $\varepsilon^{NR}_C$:
\ben p^{NR}_C=p^{NR}(0)={{m_*^2}\over{4\pi r_*^4}}\,
{{1-1/\gamma}\over{|\theta^\prime(\xi_*)|^2}} \to
\infty.\la{NRpC}\een
Hence,
\ben w^{NR}_C=w^{NR}(0)=
{{1-1/\gamma}\over{\xi_*|\theta^\prime(\xi_*)|}}\,{{m_*}\over{r_*}}
\to \infty\la{wlimit}\een
diverges only as $1/r_*$.

The corresponding limit of the coefficient $K$ depends on the
polytropic index $\gamma$: \ben K\sim
r_*^{3\gamma-4}\to\begin{cases} \infty
, &\text{if $\gamma\in [6/5,4/3)$}; \,\,\\
const\neq 0,\infty, & \text{if $\gamma=4/3;$} \,\,\\
0, & \text{if $\gamma>4/3.$}
\end{cases}\,\,\,\la{K}\een
Hence, during the above limiting process this parameter in EOS
(\ref{polytropic}) is changing if $\gamma\neq 4/3$. It is well
known that the values $\gamma< 4/3$ lead to unstable solutions,
the value $\gamma = 4/3$ corresponds to stars, build of
ultra-relativistic matter, and only the values $\gamma > 4/3$
describe stable Newtonian stars. In the limit $\gamma\to\infty$
one obtains the following results for incompressible
non-relativistic  stars:
\begin{subequations}\label{incN:xyzrt}
\ben
\theta=1-\xi^2/\xi_*^2,\, \xi_*=\sqrt{6},\la{incN:x}\\
w^{NR}(\xi)=w^{NR}_C\left(1-{{\xi^2}/{\xi_*^2}}\right),\la{incN:y}\\
\varepsilon^{NR}(r)={{3m_*}\over{4\pi r_*^3}}=const,\la{incN:z}\\
p^{NR}(r)=p^{NR}_C(1-r^2/r_*^2),\,\, p^{NR}_C={{3m_*^2}\over{8\pi
r_*^4}},\la{incN:r}\\
w^{NR}_C={1\over 2}{{m_*}\over{r_*}}.\la{incN:t} \een
\end{subequations}
Hence, as one is expecting, the point particle idealization can be
reached considering incompressible Newtonian stars, too.

Thus we see in full details how one can reach a point particle
idealization in Newton gravity. It is obvious that the resulting
point particle limit does not depend on the specific form of the
energy density distribution $\varepsilon(r)$, determined by some
fixed EOS.

\section{Point Particle Limit in GR Theory of Stars}

\subsection{Point Particle Limit in Hilbert Gauge}

According to article \cite{F0409456}, the solution of GRS problem
in HG can be formulated as a boundary problem for extended
Tolman-Openheimer-Volkov (ETOV) system of ordinary differential
equations. The very boundary is unknown. It can be fixed by the
unknown values of the luminosity variable $\rho$ at the stelar
center $C$: $\rho_C\geq 0$ and at the stelar edge:
$\rho_*>\rho_C$. For given EOS $\varepsilon=\varepsilon(p)$ the
result can be represented in the form of the relations:
\ben \varepsilon(\rho;\rho_*,\rho_C)\!=\!\begin{cases}
\varepsilon\big(p(\rho;\rho_*,\rho_C)\big), & \text{if $\rho\in
[\rho_C,\rho_*]$}; \cr 0, & \text{if $\rho\,\overline\in\,
[\rho_C,\rho_*]$},\,\,\,\,\,\,\la{epsilon}\end{cases}\een
\ben m_*:=m(\rho_*,\rho_C)=\!\int_{-\infty}^{\infty}
4\pi\varepsilon(\rho;\rho_*,\rho_C)\rho^2 d\rho,\la{m}\een
\ben m_{0*}\!:=\!m_0(\rho_*,\rho_C)\!=\!\int_{-\infty}^{\infty}
{{4\pi\varepsilon(\rho;\rho_*,\rho_C)\rho^2 d\rho
}\over{\sqrt{1\!-\!2m(\rho;\rho_*,\rho_C)/\rho}}},\la{m0}\een
\ben
R_*\!:=\!R(\rho_*,\rho_C)\!=\!\int_{\rho_c}^{\rho_*}\!{\rho\,{d\rho}
\over{\sqrt{\rho\left(\rho-2m(\rho;\rho_*,\rho_C)\right)}}}.\,\la{R}\een

The last integration is over the finite interval $[\rho_C,\rho_*]$
and the integrand is a continuous function of the variable $\rho$.
Then the function $R(\rho_*,\rho_C)$ is a continuous one with
respect to both variables $\rho_C$ and $\rho_*$. As a result the
condition $R_*\to 0$ entails simultaneously two limits:
\ben \rho_C\to \rho_0\,\,\,\hbox{and}\,\,\,
\rho_*\to\rho_0,\la{rholimits}\een
where $\rho_0\geq 0$ is some unknown limiting value, which must be
the same both for $\rho_C$ and $\rho_*$.

Let us consider this limit under the two possible extra
conditions, which are physically different:

i) Under the extra condition $m_*=m(\rho_*,\rho_C)=const$:

To use the standard procedure, which gives the point-particle
limit, in this case we construct the function
$F(\rho;\rho_*,\rho_C):={{1}\over{m_*}}4\pi\rho^2\varepsilon(\rho;\rho_*,\rho_C)$.
According to our Basic Lemma (see the Appendix), we obtain from
Eq. (\ref{m})
${1\over\epsilon}F\left({{\rho-\rho_0}\over\epsilon};\rho_*,\rho_C\right)
\rightharpoondown \delta(\rho-\rho_0)$. This gives
\ben
{{4\pi}\over\epsilon}\left({{\rho-\rho_0}\over\epsilon}\right)^2
\varepsilon\left({{\rho-\rho_0}\over\epsilon};\rho_*,\rho_C\right)
\rightharpoondown m_*\delta(\rho-\rho_0).\hskip .5truecm
\la{epsilonlim}\een

Then, to calculate in this limit the integral (\ref{m0}), we
consider the function
$\varphi(\rho)=1/\sqrt{1\!-\!2m(\rho;\rho_*,\rho_C)/\rho}$ as a
test function in the meaning of integral (\ref{wlim}). This way we
obtain from integral (\ref{m0}) the relation
$$m_{0*}={{m_*}\over{\sqrt{1\!-\!2m_*/\rho_0}}}.$$

ii). Under the extra condition $m_{0*}=m_0(\rho_*,\rho_C)=const$:

Now the point particle limit requires consideration of the
function $F(\rho;\rho_*,\rho_C):={{1}\over{m_{0*}}}
{{4\pi\varepsilon(\rho;\rho_*,\rho_C)\rho^2
}\over{\sqrt{1\!-\!2m(\rho;\rho_*,\rho_C)/\rho}}}$. Then the
application of the Basic Lemma and Eq. (\ref{m0}) gives
\ben {{4\pi}\over\epsilon}{{({{\rho-\rho_0}\over\epsilon})^2
\varepsilon({{\rho-\rho_0}\over\epsilon};\rho_*,\rho_C)
}\over{\sqrt{1\!-\!2m({{\rho-\rho_0}\over\epsilon};
\rho_*,\rho_C)/({{\rho-\rho_0}\over\epsilon})}}}\rightharpoondown
m_{0*}\delta(\rho-\rho_0).\hskip .5truecm\la{epsilonlim0}\een

To calculate the integral (\ref{m}) in the present limit, we have
to introduce in it the function
$\varphi(\rho)=\sqrt{1\!-\!2m(\rho;\rho_*,\rho_C)/\rho}$ as a test
function in the meaning if integral (\ref{wlim}). After
simultaneous multiplication and division of the integrand by the
last function $\varphi(\rho)$, and using Eq. (\ref{epsilonlim0}),
we obtain from integral (\ref{m}) the relation
$$m_{*}=m_{0*}\sqrt{1\!-\!2m_*/\rho_0}.$$

Hence, in both cases i) and ii) we arrive at the same relation,
which defines the value of $\rho_0$ via the masses $m_{*}$ and
$m_{0*}$:
\ben
    \rho_0 = 2 m_* /(1-\varrho_*^2)> \rho_{G*}.\la{rho_0S}\een
Here $\varrho_*:=m_*/m_{0*}\in (0,1)$ is the stelar mass defect
ratio. This way we prove that the limit value $\rho_0$ is greater
then the Schwarzschild radius of the star $\rho_{G*}$.

The presence of $\rho_0$ in delta functions (\ref{epsilonlim}) and
(\ref{epsilonlim0}) is a consequence of the fact that the mass ratio
$\varrho_*$ is in the open interval $(0,1)$. As a result, in point
particle limit we obtain strictly positive value of the luminosity
variable $\rho_0>\rho_{G*}>0$.

The comparison with the relation (\ref{rho_0}) and the more
complete description of gravitational field of massive point
particle in GR, recently found in \cite{F03}, shows that in both
cases we have obtained point particle limit of the GRS solutions
with arbitrary EOS. In addition, we see that $M=m_*$, and
$M_0=m_{0*}$, as one has to expect.

The physical difference between the two limiting procedures i) and
ii) is clear:

The mass $m_{0*}$ is defined by the initial amount of matter, the
star is build of. During the limiting procedure ii) this amount of
matter is preserved, but the mass $m_*$ changes, depending on EOS.

The mass $m_*$ depends on the concentrations of the matter.
Therefore it will be different for different concentrations of the
same amount of matter $m_{0*}$.

Viceversa, during the limiting procedure i) we are preserving the
same value of Keplerian mass $m_*$, changing the volume of the
contracting star. This is possible only via corresponding change
of the very amount of stelar matter, depending on EOS.

Clearly, such physically different limiting procedures are
possible only in the theory of GR stars. In Newtonian gravity the
masses $m_{0*}$ and $m_{*}$ are identical and we have a unique
limiting transition to the massive point particle idealization.

\subsection{Point Particle Limit in Basic Regular Gauge}

The point particle limit of body of finite dimension may look a
little bit tricky in HG, because the luminosity variable $\rho$ is
not a true radial variable in the problem at hand. Much more
naturel is the consideration of point particle idealization in
basic regular gauge (BRG), or in physical regular gauge (PRG)
\cite{F0409456, F03}. For simplicity we shall consider this
limiting procedure in BRG. In PRG we do not meet some new features
of the problem. The transition from BRG to PRG is a simple
fractional-linear transformation of the radial variable $r$,
preserving the place and the properties of the center of the
coordinate system $C$ at which $r=0$ in both gauges
\cite{F0409456, F03}.

In BRG the coordinate radius of the star is $r_*$ and in stelar
interior we have \cite{F0409456}:
\ben\rho(r)=\rho\left({r\over{r_*}};\rho_*,\rho_C\right),
\,\,\,\hbox{for}\,\,\,r\in[0,r_*],\,\,\,\,\, \la{rho_int}\een
with following basic properties:

i) $\rho\left(\eta;\rho_*,\rho_C\right)\in[\rho_C,\rho*]$ for
$\eta={{r\,}\over{r_*}}\in [0,1]$;

ii) $\rho\left(0;\rho_*,\rho_C\right)=\rho_C$,\,
$\rho\left(1;\rho_*,\rho_C\right)=\rho_*$;

iii) $\rho\left(1;\rho_*,\rho_C\right)=\rho_*$;

iv) $\rho\left(\eta;\rho_0,\rho_0\right)=\rho_0$ for all
$\rho_0\geq 0$ and $\eta\in [0,1]$;

v) $d\rho / \ d\eta\geq 0$.

In addition:
\begin{subequations}\label{mm0epsilon:abc}
\ben \varepsilon(r)&=&\!\begin{cases}
\varepsilon\left({r\over{r_*}};\rho_*,\rho_C\right), & \text{if
$r\in[0,r_*]$}; \cr 0\,, & \text{if
$r\,\overline\in\,[0,r_*]$},\,\,\,\,\,\,\end{cases}\hskip 1.truecm
\label{mm0epsilon:a}\\
m(r)&=&\!\begin{cases} m\left({r\over{r_*}};\rho_*,\rho_C\right),
& \text{if $r\in[0,r_*]$}; \cr m_*\,, & \text{if $r\geq
r_*$},\,\,\,\,\,\,\end{cases}\hskip 1.truecm
\label{mm0epsilon:b}\\
m_0(r)&=&\!\begin{cases}
m_0\left({r\over{r_*}};\rho_*,\rho_C\right), & \text{if
$r\in[0,r_*]$}; \cr m_{0*}\,, & \text{if $r\geq
r_*$},\,\,\,\,\,\,\end{cases}\hskip 1.truecm
\label{mm0epsilon:c}\een
\end{subequations}
and
\begin{subequations}\label{m*m0*:ab}
\ben m_*\!=\hskip 7.2truecm\label{m*m0*:a}\\
\int_{-\infty}^{\infty}\!{dr\over{r_*}}
4\pi\varepsilon \left(\!{r\over{r_*}};\rho_*,\rho_C\!\right)\!
\rho\!\left(\!{r\over{r_*}};\rho_*,\rho_C\!\right)^{\!2}\!\!
{{d\rho}\over{d\eta}}\!\left(\!{r\over{r_*}};\rho_*,\rho_C\!\right)\!,
\nonumber\\\nonumber\\
m_{0*}\!=\hskip 7.2truecm\label{m*m0*:b}\\
\!\int_{-\infty}^{\infty}\!{dr\over{r_*}} {{4\pi\varepsilon
\left(\!{r\over{r_*}};\rho_*,\rho_C\!\right)\!
\rho\!\left(\!{r\over{r_*}};\rho_*,\rho_C\!\right)^{\!2}\!\!
{{d\rho}\over{d\eta}}\!\left(\!{r\over{r_*}};\rho_*,\rho_C\!\right)
} \over{\sqrt{1\!-\!2m({r\over{r_*}};\rho_*,\rho_C)\Big/
\rho\!\left(\!{r\over{r_*}};\rho_*,\rho_C\!\right)}}}.\nonumber
\la{m*m0*}\een
\end{subequations}
 As we see, in BRG the expressions for the
corresponding quantities, although more complicated, automatically
appear in the form, suitable for study of the point particle limit
$r_*\to 0$, much like in Newtonian theory of stars, see Section
II.

Besides, we have the following two condition
\begin{subequations}\label{algebraic:ab}
\ben {m_*}^2 \int_{\rho_C}^{\rho_*}{{d\rho}\over{\rho^2}}
\sqrt{{-g_{\rho\rho}}\over{g_{tt}}}-r_*e^{\digamma}=0,\,\,\,
\la{algebraic:a}\\
\left({{m_*}\over{m_{0*}}}\right)^2\!\exp\!\left(2{r_*}\over{m_*}\right)\!+\!
{{2m_*}\over{\rho_*}}-1=0, \la{algebraic:b}\een
\end{subequations}
for matching the interior solution of EE with the exterior one in
BRG \cite{F0409456}. In Eq. (\ref{algebraic:a}) $\digamma\in
(-\infty,\infty)$ is an arbitrary fixed parameter, which describes
the stelar surface tension. This equation shows that in the limit
$r_*\to 0$ we have $\rho_*\to \rho_0$, $\rho_C\to \rho_0$. The
boundary value $\rho_0$ of the luminosity variable is determined
by the second equation (\ref{algebraic:b}), which obviously
coincides with Eq. (\ref{rho_0S}) in the limit under
consideration.

Now we are ready to consider the two physically different
limits $r_*\to 0$, which lead to point particle
idealization in BRG.

i) For fixed $m_*$:

We will obviously satisfy the requirements, described in the Note
to the Basic Lemma (see the Appendix), if we consider the function
\ben F(\eta;\rho_*,\rho_C;1,0):=\hskip 4truecm\\{{4\pi}\over{m_*}}
\varepsilon(\eta;\rho_*,\rho_C)\rho(\eta;\rho_*,\rho_C)^2
{{d\rho}\over{d\eta}}(\eta;\rho_*,\rho_C)\,\,\,\hbox{for}\,\,\eta\in[0,1],
\nonumber\la{F10}\een which is identically zero for $\eta
\,\overline\in\,[0,1]$ and corresponds to the following choice of
the parameters: $x=\eta$, $x_1=0$, $x_2=1$, $x_0=r_0$,
$y_1=\rho_*$, $y_2=\rho_C$. As a result of the last definition
$\int_{0}^{1}F(\eta;\rho_*,\rho_C;1,0)d\eta\equiv 1$ and
\ben {{4\pi}\over{r_*}} \varepsilon
\left(\!{r\over{r_*}};\rho_*,\rho_C\!\right)\!
\rho\!\left(\!{r\over{r_*}};\rho_*,\rho_C\!\right)^{\!2}\!\!
{{d\rho}\over{d\eta}}\!\left(\!{r\over{r_*}};\rho_*,\rho_C\!\right)\!
\nonumber\\ \rightharpoondown m_*\delta(r). \hskip
2.6truecm\la{deltam}\een

Using the corollary of the Basic Lemma, we obtain from Eq.
(\ref{m*m0*:b}) and Eq. (\ref{intG}):
\ben m_{0*}=m_*\int_0^1 {F(\eta;\rho_0,\rho_0;1,0){d\eta}\over
{\sqrt{1-2m(\eta;\rho_0,\rho_0)/\rho(\eta;\rho_0,\rho_0)}}}=\nonumber\\
{{m_*}\over {\sqrt{1-2m_*/\rho_0}}},\hskip 3truecm\nonumber\een
Here we have used the property iv) of the function
$\rho\left(\eta;\rho_*,\rho_C\right)$, which entails
$\rho\!\left(\eta;\rho_*,\rho_C\!\right)\to
\rho\!\left(\eta;\rho_0,\rho_0\!\right)=\rho_0$ and, in
combination with Eq. (\ref{rho_0S}), gives in the present limit
$m\left(\eta;\rho_*,\rho_C\!\right)\to
m\left(\eta;\rho_0,\rho_0\!\right)=m_*$.

ii) For fixed $m_{0*}$ :

Now we obtain
\ben {{4\pi}\over{r_*}} {{\varepsilon
\left(\!{r\over{r_*}};\rho_*,\rho_C\!\right)\!
\rho\!\left(\!{r\over{r_*}};\rho_*,\rho_C\!\right)^{\!2}\!\!
{{d\rho}\over{d\eta}}\!\left(\!{r\over{r_*}};\rho_*,\rho_C\!\right)
} \over{\sqrt{1\!-\!2m({r\over{r_*}};\rho_*,\rho_C)\Big/
\rho\!\left(\!{r\over{r_*}};\rho_*,\rho_C\!\right)}}} \nonumber\\
\rightharpoondown m_{0*}\delta(r). \hskip
2.6truecm\la{deltam0}\een
using in the Basic Lema the following proper function
\ben F(\eta;\rho_*,\rho_C;1,0):=\hskip
3.5truecm\\{{4\pi}\over{m_*}} {{\varepsilon
\left(\eta;\rho_*,\rho_C\right)
\rho\left(\eta;\rho_*,\rho_C\right)^{2}
{{d\rho}\over{d\eta}}\!\left(\eta;\rho_*,\rho_C\right)}
\over{\sqrt{1-2m(\eta;\rho_*,\rho_C)\big/
\rho\left(\eta;\rho_*,\rho_C\right)}}},
\hskip .5truecm\nonumber\\
\,\,\,\hbox{for}\,\,\eta\in[0,1],\hskip .5truecm
\nonumber\la{F10_}\een
which is identically zero for $\eta \,\overline\in\,[0,1]$.

In this case the Eq. (\ref{m*m0*:a}), in combination with relation
(\ref{rho_0S}), reproduces the limit
$m\left(\eta;\rho_*,\rho_C\!\right)\to
m\left(\eta;\rho_0,\rho_0\!\right)=m_*$. Having in mind that the
functions (\ref{mm0epsilon:abc}) are obtained from their HG
counterparts (\ref{epsilon})-(\ref{R}) by replacing the variable
$\rho$ with the function (\ref{rho_int}) \cite{F0409456} (for
example, $m(\eta;\rho_*,\rho_C):=
m^{HG}\big(\rho(\eta;\rho_*,\rho_C);\rho_*,\rho_C\!\big)$) we see
that the last result gives the limit
$m^{HG}(\rho;\rho_*,\rho_C)\to m^{HG}(\rho_0;\rho_0,\rho_0)=m_*$
in both limiting procedures at hand. This entails the limit
$\sqrt{-g_{\rho\rho}} \to 1/\sqrt{1-2m_*/\rho_0}$.

The detailed behavior of the physical parameters of the GRS under
the above limiting procedures depends on EOS. For example, for
incompressible GRS, in framework of general novel models,
developed in \cite{F0409458}, we obtain the following behavior of
the central energy density:
\ben \varepsilon_C=
{{3m_*}\over{4\pi(\rho_*^3-\rho_C^3)}}\varpropto
{{m_*}\over{4\pi\rho_0^3}}{{\rho_0}\over{\Delta\rho}} \to \infty,
\la{epsIncC}\een
when $\Delta\rho\to 0$. It diverges as $1/\Delta\rho$, if
$\rho_C>0$. This is much more weak divergency than in Newtonian
theory of stars, see Eq. (\ref{incN:z}). Only under the widely
accepted extra condition $\rho_C=0$ one obtains
$\varepsilon_C\varpropto 1/\Delta\rho^3$, i.e., a divergency of
the same strength as in Newtonian theory of stars.

To get the corresponding limit of the central pressure $p_C$ in
GRS of general type, we consider the quantity $w_C$
\cite{F0409458}:
\ben w_C={{1}\over{\sqrt{-g_{\rho\rho}}-\chi(\rho_*,\rho_C)}}-1.
\la{wC}\een

Using the previous results, we obtain the limit
\ben \chi(\rho_*,\rho_C)=4\pi\varepsilon\int_{\rho_C}^{\rho_*}
\rho\left(\sqrt{-g_{\rho\rho}}\right)^3\to\nonumber\\
{{m_*}\over{\rho_0}}\left(\sqrt{1-2m_*/\rho_0}\right)^{-3}.
\la{chi}\een
Together with Eq. (\ref{wC}) and Eq. (\ref{rho_0S}) this gives:
\ben w_C\to
{{\left(\sqrt{1\!-\!2m_*/\rho_0}\right)^3}\over{1\!-\!3m_*/\rho_0}}\!-\!1
=(1\!-\!\varrho)^2{{2\varrho-\!1}\over{3\varrho^2\!-\!1}}.\hskip
.3truecm \la{wClimit}\een

The last result entails the following consequences:

1) In the limits at hand the central pressure of incompressible
GRS of general type $p_C\to \infty$ with the same strength as
$\varepsilon_C\varpropto 1/\Delta\rho$, if $m_*/\rho_0<1/3$, 
because of the finite constant limit (\ref{wClimit}) of $w_C$.

2) For massive points, obtained as a limits of incompressible GRS,
the compactness $\varsigma:=2m_*/\rho_0< 2/3$. The mass defect
ratio is to fulfil the restriction $\varrho> 1/\sqrt{3}$.

\section{Concluding Remarks}

Thus far we were able to prove mathematically the existence of
point particle idealization in GR.

This confirms our previous results about solutions of EE with
point particle source in energy-momentum tensor \cite{F03}. There
a proper GR description of gravitational field of {\em massive}
point particle was reached for the first time.

Now we see that the gravitational field of massive point particle
in GR can be considered in mathematically and physically correct
way as a proper limiting case of the field of finite body.
Moreover, generalizing the corresponding Newtonian treatment of
this problem, we see that in GR we can consider two types of such
limits: 

i) under fixed Keplerian mass of the finite body; and 

ii) under fixed proper mass of this body. 

In both cases the limit of
the energy density is proportional to a Dirac $\delta$ function
and a mass defect of the resulting point particle is inherit from
the finite body.

The resulting {\em massive} point has a zero radius and zero
volume, but it is surrounded by a finite area. This specific
unusual property is possible only in curved space-times and
reflects the fact that in GR the energy distribution changes the
geometry of the space time.

In the above limit an infinite energy concentration at the place
of the resulting massive point emerge. Therefore one must expect
some essential deviations from the standard geometry of the 3D
balls of infinitesimal radius in an Euclidean space. In GR the
geometry around the massive point can not be the same as the
geometry around the empty one, or around a point with finite
energy density in it. This happens because in GR the energy
distribution changes the very space-time geometry by construction.
In the extreme case of infinite energy concentration the changes
of space-time geometry must be drastic. More details about this
subject the reader can find in \cite{F03}. Our present
consideration gives a more physical understanding of the results,
obtained in these articles.

\vskip .5truecm

{\em \bf Acknowledgments} \vskip .3truecm

The author is grateful to the High Energy Physics Division, ICTP,
Trieste, for the hospitality and for the nice working conditions
during his visit in the autumn of 2003. There an essential basic
ideas of present article were developed.

The author is vary grateful, too, to the JINR, Dubna, for the
financial support of the present article and for the hospitality
and good working conditions during his two three-months visits in
2003 and in 2004, when the most of the work has been done.

The author is tankful, too, to Prof. V.~Nesterenko and to unknown
referee of the first of the articles \cite{F03}, who raised the
problem of point particle limit of bodies of finite dimension in
GR, thus stimulating the development of general geometrical models
of relativistic stars and, especially, of present article.

\appendix

\section{The Basic Lemma for Sequential Approach to Dirac $\delta$-Function}
The sequential approach to mathematical distributions is well
known \cite{Gelfand}. It is suitable just for consideration of the
point particle limit in a uniform and mathematically strict way,
both in Newton gravity and in GR, starting from corresponding
models of stars of finite dimension.

In this approach the distributions, like 1D Dirac function
$\delta(x-x_0)$, are defined as a weak limits of different special
sequences of usual continuous functions $f_{\epsilon}(x,x_0)$
which, in addition, depend on some parameter $\epsilon\to 0$. We
denote by the symbol "$\rightharpoondown$" the limit in a weak
sense. This means, that when we are considering an arbitrary test
function $\varphi(r)$ (see \cite{Gelfand}) we will have
\ben \lim_{\epsilon\to 0}
\int_{-\infty}^{\infty}f_{\epsilon}(x)\varphi(x)=\varphi(x_0).
\la{wlim}\een
Then we write down
\ben f_{\epsilon}(x,x_0)\rightharpoondown
\delta(x-x_0)\,\,\,\hbox{for}\,\,\,\epsilon\to 0.
\,\la{DiracFunction}\een

We ware not able to find in the literature most general
consideration of the class of functions $f_{\epsilon}(x,x_0)$
which can be used to define the 1D Dirac $\delta$-function in the
above sense, although a lot of specific examples are well known.
Therefore we give here some general consideration, which is most
suitable for our purposes.

{\bf \em Basic Lemma:} Let us consider a real numbers $x_2>x_1$
and an arbitrary real function $F(x;x_2,x_1)$ with the following
properties:

1. $F(x;x_2,x_1)\in {\cal C}^{0}_{[x_1,x_2]}$, i.e.,
$F(x;x_2,x_1)$ is a continuous function of $x$ on the compact
interval $[x_1,x_2]$;

2.  $F(x;x_2,x_1)$ is identically zero outside this interval:
$F(x;x_2,x_1)\equiv 0\,\,\,\hbox{-- for}\,\,\,
x\,\overline\in\,\in [x_1,x_2]$;

3. $\int_{x_1}^{x_2}dx\, F(x;x_2,x_1)=1$;

Then the weak limit of the function
\ben f_{\epsilon}(x,x_0):={1\over{\epsilon}}
F\left({x-x_0\over{\epsilon}};x_2,x_1\right)\la{fF}\een
is the Dirac $\delta$-function $\delta(x-x_0)$. Here $x_0$ is an
arbitrary real number.

{\bf \em Comment:}

Note that the function $f_{\epsilon}(x,x_0)$ (\ref{fF}) is a
continuous not-identically-zero one in the interval $[x_0+\epsilon
x_1, x_0+\epsilon x_2] \to [x_0,x_0]$. At the same time its
magnitude increases to infinity, because of the factor $1/\epsilon$.

{\bf \em Proof:}

For our purposes it is enough to consider test functions
$\varphi(x)\in {\cal C}^{0}_{(-\infty,\infty)}$. Then, using the
second property of the function $F(x;x_2,x_1)$ and performing
corresponding changes of variables, we obtain:
\ben\int_{-\infty}^{\infty}dx\,f_{\epsilon}(x)\varphi(x)=\nonumber\\
\int_{x_0+\epsilon x_1}^{x_0+\epsilon x_2}dx\,
{1\over{\epsilon}}F\left({x-x_0\over{\epsilon}};x_2,x_1\right)\varphi(x)=
\nonumber\\
\int_{x_1}^{x_2}dx\,F\left(x;x_2,x_1\right)\varphi(x_0+\epsilon
x). \nonumber\een

According to first property of the function $F(x;x_2,x_1)$, when
$\epsilon \to 0$ the functions
$\psi_{\epsilon}(x;x_0,x_1,x_2):=F(x;x_2,x_1)\varphi(x_0+\epsilon
x)\in {\cal C}^{0}_{[x_1,x_2]}$  have {\em a uniform} point limit:
$\psi_{\epsilon}(x;x_0,x_1,x_2)\to F(x;x_2,x_1)\varphi(x_0)$ at
every point $x\in [x_1,x_2]$. This allows us to commute the limit
$\epsilon\to 0$ and the integration. Thus, taking into account the
third property of the function $F(x;x_2,x_1)$, we obtain:
\ben\lim_{\epsilon\to
0}\int_{-\infty}^{\infty}\!\!dx\,f_{\epsilon}(x,x_0)\varphi(x)=\nonumber\\
\int_{x_1}^{x_2}\!\!\!dx\lim_{\epsilon\to 0}
\Big(F\left(x;x_2,x_1\right)\varphi(x_0+\epsilon x)\Big)=\nonumber\\
\left(\int_{x_1}^{x_2}\!\!\!dx\,F(x;x_2,x_1)\right)\varphi(0)=
\varphi(0).\nonumber \een

{\bf \em Corollary:} For function $F(x;x_2,x_1)$ with the
properties 1-3 in the formulation of the Basic Lema, and a second
function $G(x)\in {\cal C}^{0}_{[x_1,x_2]}$, with arbitrary
properties outside the interval $[x_1,x_2]$, we have
\ben f_{\epsilon}(x,x_0)G_\epsilon(x,x_0)\rightharpoondown \,\,
<\!G\!>_F\!\delta(x-x_0)\,\,\,\hbox{for}\,\,\,\epsilon\to 0,
\hskip .3truecm \la{limFG}\een where
$f_{\epsilon}(x):={1\over{\epsilon}}F\left({x-x_0\over{\epsilon}};x_2,x_1\right)$,
$G_\epsilon(x):=G({{x-x_0}\over\epsilon})$ and
\ben <\!G\!>_F =\int_{x_1}^{x_2}\!\!dx\,
F(x;x_2,x_1)G(x).\la{intG}\een
The proof is a simple repetition of the above proof of the Basic
Lemma, after inclusion of the function $G(x)$ as a multiplier of
function $F(x;x_2,x_1)$ in all considerations.

In particular, this corollary permits us to retire the property 3
of the function $F(x;x_2,x_1)$ and to formulate a more general
result for functions without such property.

{\bf \em Note:}

One more generalization of the Basic Lemma in other direction is
needed for our purposes.

We will have the same result and the same proof, if we consider
instead of function $F(x;x_2,x_1)\in {\cal C}^{0}_{[x_1,x_2]}$ a
more general one $F(x;y_1,y_2,...;x_2,x_1)\in {\cal
C}^{0}_{[x_1,x_2]\times \mathbb{R}_{y_1}\times
\mathbb{R}_{y_2}\times...}$, i.e., a continuous function of many
variables $x,y_1,y_2,...$, and suppose that in it
$y_1=y_1(\epsilon)$, $y_2=y_2(\epsilon)$, ... are ${\cal
C}^{0}_{\mathbb{R}}$ functions of the variable $\epsilon$ with
definite limits $y_1(0)$, $y_2(0)$, ..., when $\epsilon\to 0$. In
addition, it is supposed that the function
$F(x;y_1(\epsilon),y_2(\epsilon),...;x_2,x_1)$ owns the properties
2 and 3 for any small enough $\epsilon$.

\end{document}